\documentclass[conference]{IEEEtran}
\usepackage[colorinlistoftodos, textwidth=2cm,textsize=tiny]{todonotes}
\usepackage{xspace}
\usepackage{graphicx}
\usepackage{paralist}

\usepackage{algorithmicx}
\usepackage{algorithm}
\usepackage[noend]{algpseudocode}
\usepackage{amsmath}
\usepackage{amssymb}
\usepackage{multirow}
\usepackage{booktabs}
\usepackage[normalem]{ulem}

\usepackage{balance} % To use \balance in the end

\usepackage[compress, sort]{cite} % transforms [1][2][3] into [1]-[3]

\usepackage{graphicx}
\usepackage{float}
\usepackage[caption = false]{subfig}

\usepackage{url}

\usepackage[export]{adjust box}
\usepackage{pgf}
\begin{document}
\frenchspacing
\setlength{\pdfpagewidth}{8.5in}
\setlength{\pdfpageheight}{11in}
% % Paragraphs
\newcommand{\rahul}[1]{\textcolor{blue}{\textbf{\tiny \textless rahul add\textgreater }#1}}
\newcommand{\rahulst}[1]{\textcolor{red}{\textbf{\tiny{\textless rahul delete\textgreater }}\sout{#1}}}
\newcommand{\ganesh}[1]{\textcolor{purple}{\textbf{\tiny \textless ganesh add\textgreater }#1}}
\newcommand{\ganeshst}[1]{\textcolor{red}{\textbf{\tiny{\textless ganesh delete\textgreater }}\sout{#1}}}

\newcommand{\spara}[1]{\smallskip\noindent{\bf #1}}
\newcommand{\mpara}[1]{\medskip\noindent{\bf #1}}
\newcommand{\para}[1]{\noindent{\bf #1}}

\title{
Relevancy Classification of Multimodal Social Media Streams for Emergency Services}  

\author{\IEEEauthorblockN{Ganesh Nalluru}
\IEEEauthorblockA{\textit{George Mason University}\\
Fairfax, VA, USA \\
gn@gmu.edu}
\and 
\IEEEauthorblockN{Rahul Pandey}
\IEEEauthorblockA{\textit{George Mason University}\\
Fairfax, VA, USA \\
rpandey4@gmu.edu}
\and 
\IEEEauthorblockN{Hemant Purohit}
\IEEEauthorblockA{\textit{George Mason University}\\
Fairfax, VA, USA \\
hpurohit@gmu.edu} 
}

\maketitle

\begin{abstract}
Social media has become an integral part of our daily lives. During time-critical events, the public shares a variety of posts on social media including reports for resource needs, damages, and help offerings for the affected community. Such posts can be relevant and may contain valuable situational awareness information. However, the information overload of social media challenges the timely processing and extraction of relevant information by the emergency services. Furthermore, the growing usage of multimedia content in the social media posts in recent years further adds to the challenge in timely mining relevant information from social media.  

In this paper, we present a novel method for multimodal relevancy classification of social media posts, where relevancy is defined with respect to the information needs of emergency management agencies. Specifically, we experiment with the combination of semantic textual features with the image features to efficiently classify a relevant multimodal social media post.  
We validate our method using an evaluation of classifying the data from three real-world crisis events.  
Our experiments demonstrate that features based on the proposed hybrid framework of exploiting both textual and image content improve the performance of identifying relevant posts. 
In the light of these experiments, the application of the proposed classification method could reduce cognitive load on emergency services, in filtering multimodal public posts at large scale. 
\end{abstract}


%%%%%%%%%%%%%%%%%%%%%%%%%%%%%%%%%%%%%%%%%%%%%%%%%%%%%%%%
% Main content
%%%%%%%%%%%%%%%%%%%%%%%%%%%%%%%%%%%%%%%%%%%%%%%%%%%%%%%%

%%%%%%%%%%%%%%%%%%%%%%%%%%%%%%%%%%%%%%%%%%%%%%%%%%%%%%%%
\section{Introduction}
\label{sec:intro}
%%%%%%%%%%%%%%%%%%%%%%%%%%%%%%%%%%%%%%%%%%%%%%%%%%%%%%%%

Social media plays a key role during disaster events, as evident from the events in  recent years. People share all kinds of information, ranging from jokes and prayers to damage in the affected regions~\cite{imran2015processing} to requests and offers to help~\cite{purohit2013emergency}. Studies in  recent years have also shown the public expectation for timely seeking a response to their calls for help on social media~\cite{reuter2017towards}. 

For emergency management agencies, getting relevant situational awareness information from the affected public is of utmost importance. In particular, the Public Information Officers (PIOs) monitors the information from public to provide intelligence to the decision makers in the response coordination\cite{hughes2012evolving}. Given the increasing importance of social media data, emergency services have started to monitor social media~\cite{dhs2014using}. However, given their limited human resources and the vast amounts of social media messages posted with high velocity during disaster events, a critical challenge is to address the high information overload on the emergency service personnel~\cite{castillo2016big}.   
There is a recognition of the necessity to effectively filter, prioritize, and organize information from this unconventional information channel~\cite{dhs2014using}. 

Furthermore, social media messages are increasingly becoming multimodal in nature~\cite{alam2018crisismmd}. Thus, the informative data may be of different modalities other than text such as images, videos, and audios. The inclusion of data of other modalities highly affects the complexity of mining the informative messages that might be important to capture. Hence, exploiting features from all the modalities in the data can be valuable and necessary to efficiently understand the actual meaning of the message. Content of such posts vary in potential value for operational response, ranging from actionable and serviceable requests~\cite{purohit2018social} to potential rumors~\cite{starbird2014rumors} to damage reports~\cite{nguyen2017damage}. 
Thus, quickly filtering and prioritizing  
such multimodal messages with relevant information have become a critical need for response agencies~\cite{dhs2014using}. % 
\begin{figure} 
\centering
\includegraphics[width = \columnwidth]{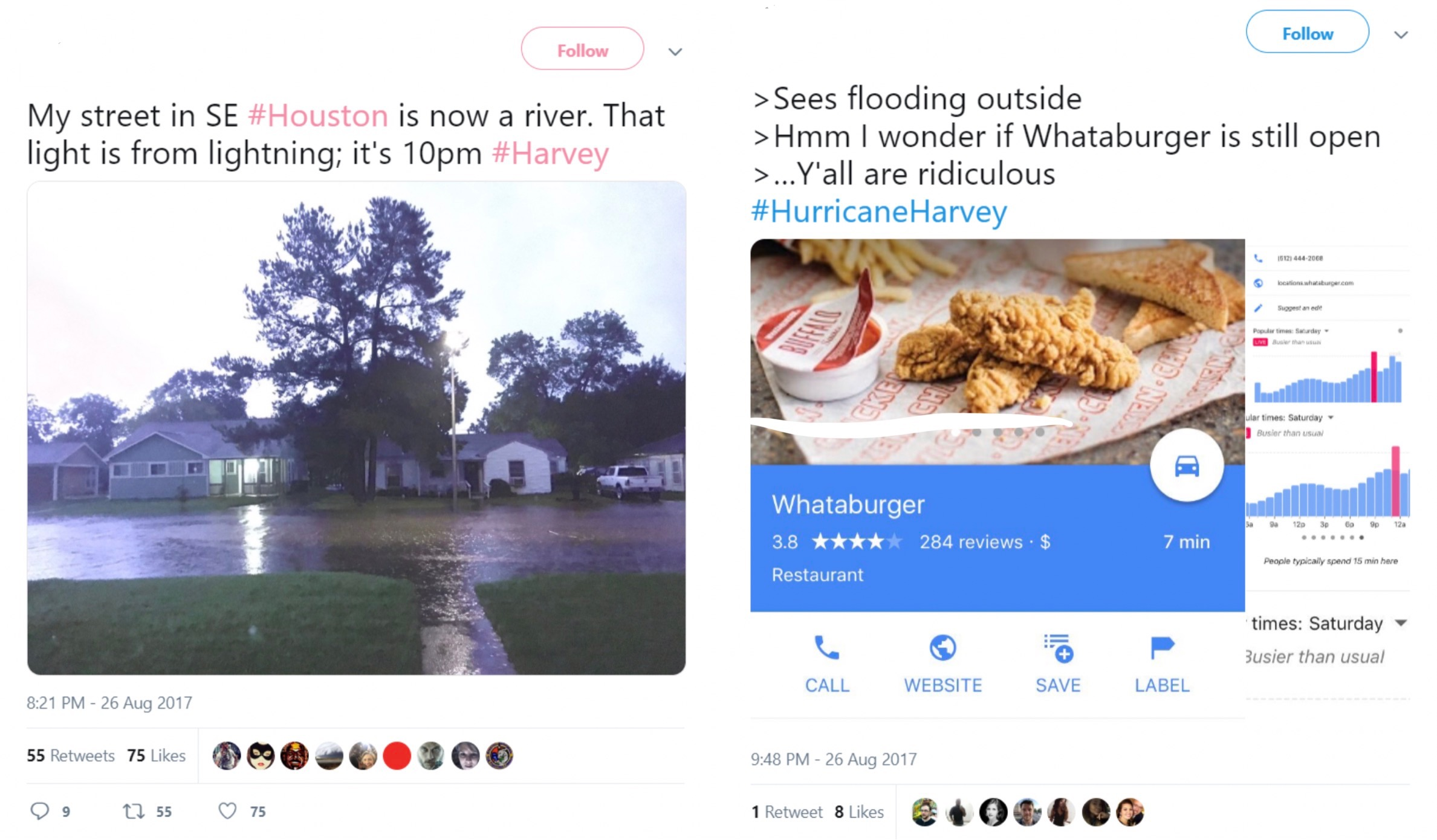}
\caption{Exemplar messages with different degree of relevance for textual and image characteristics, which provide informative vs non-informative signal to judge relevancy for the public post. \textit{(User name and id removed for anonymity)} }  
\label{fig:example}
\end{figure}

Figure~\ref{fig:example} shows some example social media messages where the embedded image content can present valuable signal to quickly judge the relevancy for public social media posts for emergency management in contrast to the irrelevant information shared during emergency events.     

\spara{Problem.}  
We address the problem of classifying relevant multimodal social media posts for helping emergency services quickly filter social media streams. 

\spara{Our contribution.} 
We provide a generalizable multimodal classification framework and present extensive experimentation to study the interplay of textual and image features to classify relevant multimodal social media posts for emergency services.   
We introduce the framework in Section~\ref{sec:approach} together with a description of textual and image features, which leverages the popular distributional semantic representations of text and image data. Finally, we demonstrate the validity of the proposed method in Section~\ref{sec:experiments}, by experimenting with the real-world Twitter datasets collected during three crisis events in the recent years. We conclude with lessons and future work directions in Section~\ref{sec:conclusion}. 

%%%%%%%%%%%%%%%%%%%%%%%%%%%%%%%%%%%%%%%%%%%%%%%%%%%%%%%%
\section{Related Work}
\label{sec:related}
%%%%%%%%%%%%%%%%%%%%%%%%%%%%%%%%%%%%%%%%%%%%%%%%%%%%%%%%
\begin{figure*} 
\centering
\includegraphics[width=7in]{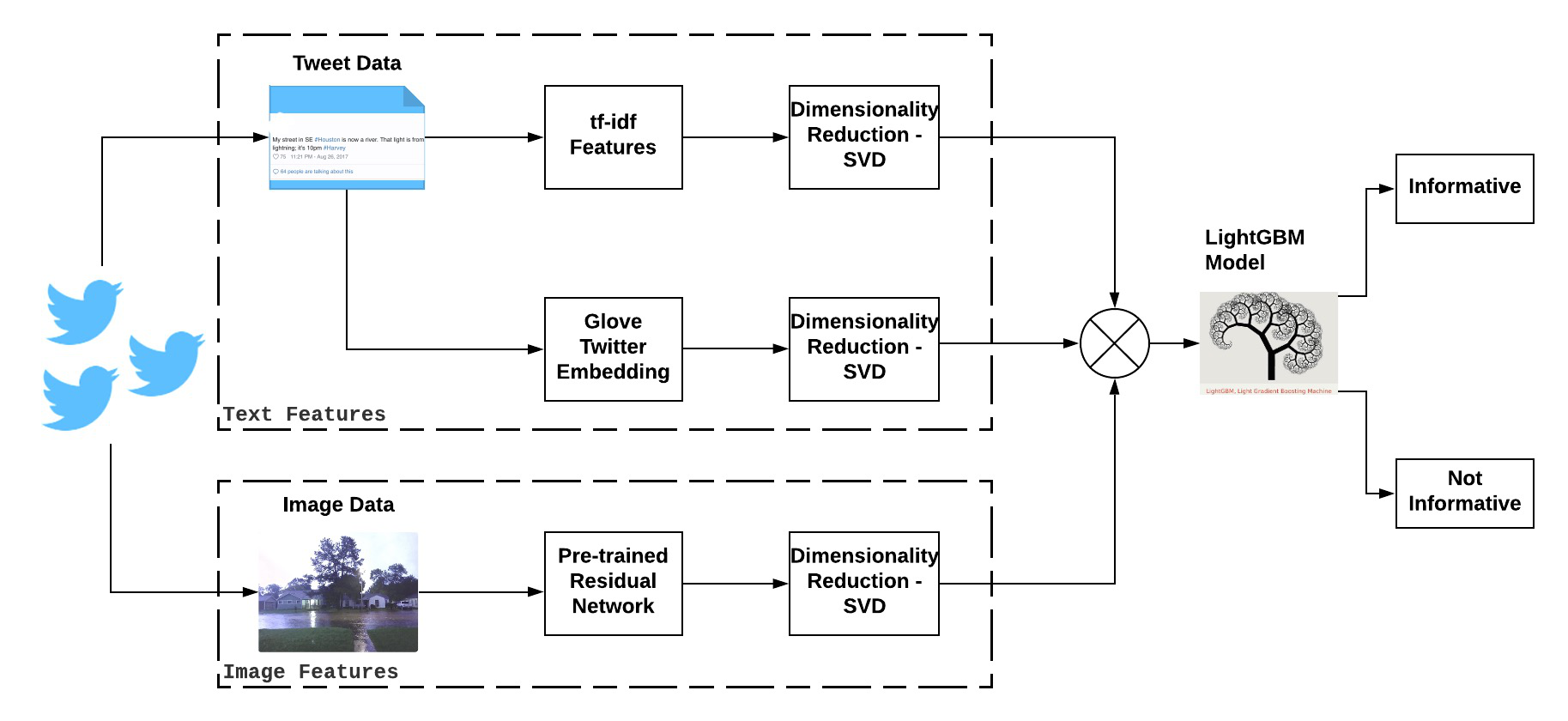}
\caption{Proposed multimodal classification framework that exploits both text and image content features trained on LightGBM model.}  
\label{fig:correlation}
\end{figure*}
There has been extensive research on the topic of social media for emergency management in the last decade~\cite{imran2015processing,castillo2016big}. The nature of data generated over social media has such a high volume, variety, and velocity causing the challenges of ``Big Crisis Data'' that often overwhelm the emergency services~\cite{castillo2016big}. The literature in crisis informatics~\cite{palen2016crisis} field has investigated social media for emergency services using diverse multidisciplinary perspectives. % of computing, information, and social sciences.   

%The emergency services have Public Information Officers (PIOs) as the critical actors who monitor social media as well as serve information to the public for helping the operational response of agencies~\cite{dhs2014using}. 
User studies with emergency responders have identified information overload as one the key barrier for efficiently using social media platforms by PIOs and emergency services~\cite{hiltz2014use,castillo2016big}. Such information overload factors include the processing of unstructured and noisy nature of multimodal social media content at large scale, which is beyond the capacity of the limited human resources. Furthermore, characterizing the relevancy of social media content is very contextual,  time-sensitive, and often challenging~\cite{purohit2018social}.   

Among the social media analytics approaches, researchers have modeled public behavior in specific emergencies, addressed the problems of data collection and filtering, classification and summarization as well as visualization of analyzed data for decision support~\cite{imran2015processing}. However, the focus of such works has centered around text analytics except recent studies~\cite{alam2018twitter,alam2018crisismmd,nguyen2017damage,basnyat2017analyzing} on processing multimedia content of the social posts. Although current multimodal information processing approaches for social media mining during disasters primarily analyzed only the damage assessment aspect of emergency management.   
We further complement these recent studies by proposing a generic classification framework for relevant information that exploits both textual and image content of multimodal social media posts.  

%%%%%%%%%%%%%%%%%%%%%%%%%%%%%%%%%%%%%%%%%%%%%%%%%%%%%%%%
\section{Method}\label{sec:approach}
\subsection{Task Definition}
Users in Twitter often share both text as well as image content during natural disasters. The goal of this task is to classify a tweet, if the tweet posted during the natural disaster is informative or not. The informativeness of a tweet is defined by the relevancy of embedded information to various emergency services or humanitarian organizations as described in ~\cite{alam2018crisismmd}, such as the message containing rescue related and affected individuals related information. 

\subsection{Summary of the Proposed Approach}
Learning features for classification of short text messages such as tweet is challenging due to limited availability of context to exploit. Often, these tweets require some context representation to extract the meaning of the text effectively. One possible solution is to use deep learning model for feature generation and text classification. However, not having the richness of large dataset results in overfitting of deep models as the features generated during the training are not generalizable enough %linear towards the training data because
 due to the large number of variables to train. Therefore, we propose a feature generation framework that utilizes both the text features and image features to enrich the context representation from the given multimodal social media posts. We then perform the dimensionality reduction on three diverse extracted feature vectors and concatenate these vectors. Finally, we build a LightGBM model~\cite{ke2017lightgbm} for multimodal relevancy classification. 

\subsection{Extracting Features}
In this section, we describe the different methods used to derive the text features and image features. % from multimodal data in the final ensemble.

\subsubsection{Text features}
We have utilized two different types of text features: \textit{Bag of Words} features and \textit{Word embeddings} features. Bag of Words features such as $tf$-$idf$ vector representation of text give high weights to a word that often appears in particular tweet, but not as frequently in the whole collection of messages. Through $tf$-$idf$, we try to capture the semantic significance of that word to the document (tweet) in our corpus.

The following steps describe our approach to compute $tf-idf$ vectors for each document/tweet. 
\begin{enumerate}[(i)]
    \item First, we create a vocabulary set of size $V$ for the entire training tweet set that contains a set of all the words present in all the tweets. 
    \item Now for each tweet, we create a vector of size $V$. Each position of the vector represents a vocabulary word and each value represents the $tf-idf$ value for that word in the tweet.
    \item To calculate $tf-idf$ value for a word $w$ in a tweet, we compute $tf(w) = D_w / N$ where $D_w$ is the number of times the word $w$ appears in that tweet and we compute $idf(w) = \log{(N/D^{'}_w)}$, where $D^{'}_w$ is the total number of tweets which contain the word $w$. Finally, the $tf-idf$ score is the multiplication of $tf(w)$ and $idf(w)$. 
\end{enumerate}

\textit{Word embeddings} provide the semantic representation of words in low-dimensional space, where a dense, real-valued vector represents each word in a fixed dimensional space such that the words with similar meaning are closer to each other. We took the average of the embedding vectors of all the words in a tweet to describe the particular tweet. Since our data is small, we  utilize the pre-trained GloVe vector embeddings~\cite{pennington2014glove} of 200 dimensions trained on Twitter data (27 billion tweets) to get the embedding vectors of each word.

Given that Bag-of-Words model with tf-idf representation does not capture the latent semantics of the tweet, thus, we combined both the Bag-of-Words with tf-idf features and Glove-Twitter word Embeddings. Such feature combination captures both the syntactic and semantic aspects of the tweet, which is important for classifying the informativeness of the tweet.

The following steps describe our approach to compute word embedding vectors for each tweet: 
\begin{enumerate}[(i)]
    \item For each tweet, we initialize a vector $V$ with all zeros of size $D$ where $D$ is the dimension size for pretrained word embeddings. 
    \item Now for each word in the tweet, we will check if we have the word embedding in our pretrained model. If it is present, we will perform the vector addition with the initial vector $V$. 
    \item Finally, we will divide the final vector $V$ with a scalar value $n$ where $n$ is the size of the subset of words in the tweet that have the embedding representation from the pre-trained model. 
\end{enumerate} 

\subsubsection{Image features}
For generating \textit{image features}, we %employ Deep Neural Networks. The reason behind using Deep Learning is because 
have to address the sparsity issue due to a diverse range of images that are highly distributed in terms of embedded information. Hence, we utilized the pre-trained model of ResNet (Residual Learning Framework)~\cite{he2016deep} of 50 layer architecture to get the features for each image. 

The following steps describe our approach to compute the image features for a given tweet:  
\begin{enumerate}[(i)]
    \item First we set up the ResNet-50 architecture and initialize it with the pretrained model weights. 
    \item We then pass each image in our training set as an input to this deep network. 
    \item We store the second last layer vector for each of the image in the input training instance tweet separately. 
    \item We use this vector as a \textit{image features} for that training instance tweet.
\end{enumerate}

\subsubsection{Feature Engineering}
As per our initial exploratory analysis on the dataset, we found few discriminative factors to differentiate \textit{informative} vs \textit{non-informative} tweets. First, we observed that all the tweets with \textit{\#news} were categorized as informative as compared to non-informative. Similarly, we observed that the length of the \textit{informative} tweets were comparatively larger than that of \textit{non-informative} tweets. As a result, we have included \textit{Word Count} and \textit{Character Count} as another set of features in our \textit{text features} category.

\subsubsection{Combining Text Features and Image Features}
A key challenge with multimodal relevancy classification task is to find a systematic and effective way to combine the text features and image features. Also, the feature dimensions are different across the two modalities. Therefore, we used a Singular Valued Decomposition (SVD) to perform dimensionality reduction on all three types of extracted feature vectors and then, concatenated these vectors for the final feature representation for training the classification model. 

\subsection{Multimodal Disaster Data Classification} 
We have specifically used LightGBM (LGBM) \cite{ke2017lightgbm} as our model for classification as it performed well on multimodal data in our preliminary testing with other models. Also, we found it be very useful particularly for time-sensitive emergency services after observing its faster performance in comparison to deep learning models. LGBM is an implementation of fast gradient boosting on decision trees. It is faster as it removes a significant proportion of data instances with small gradients, and includes only the rest to compute the information gain and also, bundle mutually exclusive features to reduce the number of features. For our baseline schemes, we have compared our proposed model with two other models i.e. Logistic Regression and XG Boost\cite{chen2016xgboost}.

\section{Experiments and Results}\label{sec:experiments}
\begin{table*}
  \caption{Comparison of Accuracy and AUC for various schemes on the 20\% test set.
  % The results from the combination of both text and image features trained on LightGBM outperform other schemes across all disaster events.
  \textit{\scriptsize REVISED NOTE: These results are different from the version hosted at IEEE Xplore with an error of inconsistent parameters in Table 1.  Results in this version are generated keeping the same parameter overall; Our conclusion still remains the same.}}
  \vspace{-0.1cm}
  \label{tab:result} 
  \centering
  \begin{tabular}{|p{1.2cm}|p{7.0cm}|r|r|r|r|r|r|r|r|} 
    \hline  
    \multirow{2}{*}{\textbf{Dataset}} & \multirow{2}{*}{\textbf{Input Features}} & 
    \multicolumn{2}{|c|}{\begin{minipage}{2.0cm} \bf  [\textit{M1}] LR \end{minipage}} &
    \multicolumn{2}{|c|}{\begin{minipage}{2.0cm} \bf  [\textit{M2}] XG Boost \end{minipage}} &
    \multicolumn{2}{|c|}{\begin{minipage}{2.0cm} \bf  [\textit{M3}] LightGBM \end{minipage}}\\ \cline{3-8}
    % &
    % \begin{minipage}{1.3cm} \bf Sufficiently Detailed \end{minipage} \\ \hline
   & & \bf Accuracy & \bf AUC & \bf Accuracy & \bf AUC & \bf Accuracy & \bf AUC \\ \hline
   \multirow{5}{1.2cm}{\textbf{Hurricane Maria}} & \textbf{[\textit{T1}] Text features - Bag of Words}   &  78.9\%  & 76.05\% & 77.3\% & 74.95\% & 76.5\% & 81.45\%\\ \cline{2-8}
   & \textbf{[\textit{T2}] Text features - tf-tdf }   &  78.5\%  & 74.27\% & 77.4\% & 74.97\% & 76.5\% & 81.34\%\\ \cline{2-8}
   & \textbf{[\textit{T3}] Text features - (tf-idf + GloVe Embeddings)}   &  77\%  & 73.45\% & 80.1\% & 77.01\% & 81.1\% & 88.06\%\\ \cline{2-8}
   & \textbf{[\textit{T2+I1}] Text features - tf-idf + Image features - ResNet}   
   &  75.4\%  & 71.5\% & 77.77\% & 74.38\% & \bf 78.8\% & \bf 84.7\%\\ \cline{2-8}
   & \textbf{[\textit{T3+I1}] Text features - (tf-idf + GloVe Embeddings) + Image features - ResNet}   &  77.3\%  & 74.02\% & 81.3\% & 78.04\% & \bf 81.3\% & \bf 88.18\%\\ \hline
   \multirow{5}{1.2cm}{\textbf{Hurricane Harvey}} & \textbf{[\textit{T1}] Text features - Bag of Words}   &  85.3\%  & 74.3\% & 85\% & 75.2\% & 82.4\% & 86.13\%\\ \cline{2-8}
   & \textbf{[\textit{T2}] Text features - tf-idf }   &  80.6\%  & 60.6\% & 84.6\% & 74.7\% & 82.7\% & 85\%\\ \cline{2-8}
   & \textbf{[\textit{T3}] Text features - (tf-idf + GloVe Embeddings)}   &  83.9\%  & 72.99\% & 86.9\% & 77.97\% & 86.9\% & 90.0\%\\ \cline{2-8}
   & \textbf{[\textit{T2+I1}] Text features - tf-idf + Image features - ResNet}   &  79.2\%  & 61.55\% & 83.1\% & 70.39\% & \bf 82.6\% & \bf 85.99\%\\ \cline{2-8}
   & \textbf{[\textit{T3+I1}] Text features - (tf-idf + GloVe Embeddings) + Image features - ResNet}   &   83.9\%  & 73.66\% & 86.4\% & 76.80\% & \bf 87.4\% & \bf
   90.65\%\\ \hline
   \multirow{5}{1.2cm}{\textbf{Hurricane Irma}} & \textbf{[\textit{T1}] Text features - Bag of Words}   &  82.8\%  & 65.33\% & 83.1\% & 65.36\% & 81.8\% & 82.15\%\\ \cline{2-8}
   & \textbf{[\textit{T2}] Text features - tf-idf }   &  82.1\%  & 59.11\% & 83.3\% & 66.63\% & 82.4\% & 80.46\%\\ \cline{2-8}
   & \textbf{[\textit{T3}] Text features - (tf-idf + GloVe Embeddings)}   &   83.5\% & 66.58\% & 85.2\% & 70.03\% & 85.3\% & 87.20\%\\ \cline{2-8}
   & \textbf{[\textit{T2+I1}] Text features - tf-idf + Image features - ResNet}   &  80.9\%  & 56.56\% & 83.6\% & 66.46\% & \bf 83.4\% & \bf 83.13\%\\ \cline{2-8}
   & \textbf{[\textit{T3+I1}] Text features - (tf-idf + GloVe Embeddings) + Image features - ResNet}   &  84.0\%  & 67.28\% & 84.9\% & 69.08\% & \bf 85.4\% & \bf 88.01\%\\ \hline
\end{tabular}
\vspace{-0.25cm}
\end{table*}
\subsection{Dataset}
We conduct our experiments on three large disaster events from the year 2017 - Hurricane Maria, Hurricane Harvey, and Hurricane Irma, where the data was provided by~\cite{alam2018crisismmd}. The dataset contains Twitter posts with both text and their associated images as well as class label: \textit{informative} or \textit{non-informative}. The total \textit{informative} and \textit{non-informative} tweets across the events were 3095 and 984 for Harvey respectively, 3320 and 889 for Irma, and 2569 and 1528 for Maria respectively. %, in the context of relevancy for emergency management. 
We performed standard text preprocessing on tweets, including removing URLs, special symbols such as `RT @user', stop words, and lowercasing. 

\subsection{Classification Schemes}
We used 80-20 split strategy to create train-test sets and the AUC (Area Under the ROC Curve) and Accuracy measures to evaluate performance on the 20\% test. We employ various classification schemes for training with different features and learning algorithms as follows (\textit{Mx} denotes model type, \textit{Tx} text feature type, and \textit{Ix} image feature): % Following are the schemes we have used for training.
\begin{itemize}
    \item \textbf{[\textit{T1 + M1}] Text features (Bag of words) + Logistic Regression Model} (baseline) This method uses Bag-of-Words representation to generate text features (T1) and train that on Logistic Regression model.
    \item \textbf{[\textit{T1 + M2}] Text features (Bag of words) + XGBoost Model} This method uses same features of T1 and train that on XGBoost model.
    \item \textbf{[\textit{T1 + M3}] Text features (Bag of words) + LGBM Model} This method uses same features of T1 and train that on LGBM model.
    \item \textbf{[\textit{T2 + M1}] Text features (tf-idf) + Logistic Regression Model} This method uses tf-idf to generate text features (T2) and train that on Logistic Regression model.
    \item \textbf{[\textit{T2 + M2}] Text features (tf-idf) + XGBoost Model} This method uses same features of T2 and train that on XGBoost model. 
    \item \textbf{[\textit{T2 + M3}] Text features (tf-idf) + LGBM Model} This method uses same features of T2 and train that on LGBM model. 
    \item \textbf{[\textit{T3 + M1}] Text features (tf-idf + GloVe Embeddings)  + Logistic Regression Model}  This method uses the concatenation of both tf-idf and GloVe Twitter embeddings to generate text features (T3) and train that on Logistic Regression model.
    \item \textbf{[\textit{T3 + M2}] Text features (tf-idf + GloVe Embeddings) + XGBoost Model} This method uses same features of T3 and train that on XGBoost model.
    \item \textbf{[\textit{T3 + M3}] Text features (tf-idf + GloVe Embeddings) + LightGBM Model} This method uses same features of T3 and train that on LGBM model.
    \item \textbf{[\textit{T2 + I1 + M1}] Text features (tf-idf) + Image Features (ResNet) + Logistic Regression Model} This method uses the concatenation of both Tf-Idf as text features (T2) and ResNet features as Image features (I1) and train that on Logistic Regression model. 
    \item \textbf{[\textit{T2 + I1 + M2}] Text features (tf-idf) + Image Features (ResNet) + XGBoost Model} This method uses same features of T2 and I1, and train that on XGBoost model.
    \item \textbf{[\textit{T2 + I1 + M3}] Text features (tf-idf) + Image Features (ResNet) + LGBM Model} This method uses same features of T2 and I1, and train that on LGBM model.
    \item \textbf{[\textit{T3 + I1 + M1}] Text features (tf-idf + GloVe Embeddings) + Image Features (ResNet) + Logistic Regression Model} This method concatenates tf-Idf and GloVe Embeddings as text features and ResNet features as Image features, and train a Logistic Regression model.
    \item \textbf{[\textit{T3 + I1 + M2}] Text features (tf-idf + GloVe Embeddings) + Image Features (ResNet) + XGBoost Model} This method uses same features of T3 and I1, and train that on XGBoost model.
    \item \textbf{[\textit{T3 + I1 + M3}] Text features (tf-idf + GloVe Embeddings) + Image Features (ResNet) + LGBM Model} (proposed) This method uses same features of T3 and I1, and train that on LGBM model. 
\end{itemize}

\subsection{Results}
Table \ref{tab:result} shows the results of our proposed features with LGBM model compared to the other baselines (XGBoost and Logistic Regression). The table shows the performance of different feature representations evaluated based on accuracy score and AUC. 
Our proposed approach is using both types of text features i.e. $tf-idf$ and word embeddings as well as image features and trained on Light Gradient Boosting (LGBM). We have reported the results for three event datasets and analyzed how our proposed approach is performing compared to other baselines.

\subsection{Discussions}\label{sec:discussions}
We built several machine learning models based on our hybrid feature representation (combining $tf-idf$, Glove Embedding and Resnet features) and compared against the baseline of text-only features. 

Table \ref{tab:result} shows that our proposed method with LightGBM Model outperformed other models in all the experiments on our hybrid feature representation of text and image features, across all disaster event datasets. 
We also noticed that the scheme of Logistic Regression model performed poorly when combining both $tf-idf$ features (with an AUC of $74.2\%$ in Hurricane Maria event) and image features (Resnet-50 features), giving us an AUC of $71.5\%$ (similar case for other disaster events as well). It is likely due to simpler learning algorithm. 
In contrast, XGBoost Model performance on both $tf-idf$ features and our proposed hybrid features respectively gave us almost the same performance. XGBoost Model with $tf-idf$ features gave us an AUC of $74.97\%$ in Hurricane Maria and AUC was nearly the same when we added image features (Resnet-50). 

Our proposed method performs better as it combines both the $tf-idf$ features and word embedding features, which capture both the syntactic and semantic aspects from text. Adding image features improves the performance and helps the model to learn from multimodal features. Based on these results, we can conclude that the addition of image features (ResNet-50) to text features improves the performance in classifying \textit{informative} tweets.

\section{Conclusions}\label{sec:conclusion}
%%%%%%%%%%%%%%%%%%%%%%%%%%%%%%%%%%%%%%%%%%%%%%%%%%%%%%%%

This paper presented a novel approach of classifying relevant multimodal social media posts by systematically combining textual and image features, which are able to capture latent semantics of relevant information within the short-text messages. We have validated our approach with three crisis event datasets and achieved over 90\% AUC score. The proposed method can help improve social media services at emergency response organizations. 
We plan to further evaluate our framework across different social media platforms for generalizability. 
Furthermore, there is a vast literature in the computer vision community on caption generation approaches for the textual description of images, %. Our future work will also 
and we plan to leverage these approaches to extract the textual information from the images for getting additional semantics. % in the presented hybrid feature framework. %using Image captioning and then c the presented feature tfidf and glove embedding features.

\balance

%%%%%%%%%%%%%%%%%%%%%%%%%%%%%%%%%%%%%%%%%%%%%%%%%%%%%%%%
\section{Acknowledgement}
Authors thank US National Science Foundation (NSF) grant IIS-1657379 for partial support to this research. %Opinions in this paper are those of the authors and do not necessarily represent the social position or policies of the NSF.   
%
%%%%%%%%%%%%%%%%%%%%%%%%%%%%%%%%%%%%%%%%%%%%%%%%%%%%%%%%

\bibliographystyle{IEEEtran} 
\bibliography{paper-smartsys19}

\end{document}